\documentclass[manuscript]{acmart}
\usepackage{graphicx} 
\usepackage{subcaption}
\usepackage{float}
\usepackage{amsmath}

\title{Quantifying Media Influence on Covid-19 Mask-Wearing Beliefs}
\author{Nicholas Rabb, Nitya Nadgir, Jan P. de Ruiter, Lenore Cowen}
\date{August 2023}

\begin{document}

\begin{abstract}
    How political beliefs change in accordance with media exposure is a complicated matter. 
    Some studies have been able to demonstrate that groups with different media diets in the aggregate (e.g., U.S. media consumers ingesting partisan news) arrive at different beliefs about policy issues, but proving this from data at a granular level -- at the level of attitudes expressed in news stories -- remains difficult. In contrast to existing opinion formation models that describe granular detail but are not data-driven, or data-driven studies that rely on simple keyword detection and miss linguistic nuances, being able to identify complicated attitudes in news text and use this data to drive models would enable more nuanced empirical study of opinion formation from media messaging.
    This study seeks to contribute a dataset as well as an analysis that allows the mapping of attitudes from individual news stories to aggregate changes of opinion over time. Additionally, the subject of our study is an important public health topic where opinion differed in the U.S. by partisan media diet: Covid mask-wearing beliefs.
    By gathering a dataset of U.S. news media stories, from April 6 to June 8, 2020, annotated according to Howard 2020's Face Mask Perception Scale for their statements regarding Covid-19 mask-wearing, we demonstrate fine-grained correlations between media messaging and empirical opinion polling data from a Gallup survey conducted during the same period. We also demonstrate that the data can be used for quantitative analysis of pro- and anti-mask sentiment throughout the period, identifying major events that drove opinion changes. 
    This dataset is made publicly available and can be used by other researchers seeking to evaluate how mask-wearing attitudes were driven by news media content. Additionally, we hope that its general method can be used to enable other media researchers to conduct more detailed analyses of media effects on opinion.
\end{abstract}

\maketitle

\section{Introduction}

During the early Covid-19 pandemic, survey and polling studies indicated that important public health measures, such as wearing a face mask, were being adhered to at different rates based on individuals' political partisan identity \cite{Clinton2021PartisanCOVID-19,Kahane2021PoliticizingUSA,Gramlich2020202020,Newport2020TheCoronavirus}. Consistently, those identifying as Republicans were shown to mask less often than those identifying as Democrats, judge the virus as less dangerous, and cast more doubt on social distancing and other measures. Indeed, even by February, 2021, almost a year after the declaration of the pandemic by the World Health Organization, though Pew Research reported that the gap between mask-wearing behavior had significantly shrunk (in June, 2020, while 76\% of Democrats reported regularly wearing masks in stores and businesses, only 53\% of Republicans reported the same; but in February, 2021, the numbers shifted to 90\% of Democrats as compared to 83\% of Republicans) \cite{Schaeffer2021DespiteExists}, significant gaps still existed in other public health measures such as requiring face masks on airplanes or public transportation (96\% of Democrats saying this is necessary compared to 72\% of Republicans) or avoiding gatherings in large groups (93\% of Democrats saying this is necessary compared to 56\% of Republicans). Partisan differences in perceptions towards, and adherence to, public health measures persisted into January, 2022, according to follow-up polling that reported significant partisan gaps in whether or not respondents thought the coronavirus outbreak had been made a bigger deal than it actually was (16\% of Democrats versus 64\% of Republicans) \cite{Mitchell2022AttentionOutbreak} and in vaccination rates (90\% of Democrats versus 64\% of Republicans) \cite{Tyson2022IncreasingU.S.}.

To explain this discrepancy in belief in, and adherence to, public health measures, some commentators argued that different media diets may have been driving different beliefs and subsequent behaviors \cite{Bird2020IsPractices,Jurkowitz2020CableSource}. While partisan differences in consumption of political news (i.e. that U.S. media has become more partisan in nature, and subsequently partisans increasingly turn to different sources for their news in both traditional and social media) is a well-studied phenomenon \cite{Iyengar2009RedUse,Bakshy2015ExposureFacebook}, the fact that public health information became something that was subject to partisan differences in reporting -- to the degree that partisan media diets may have been driving wide gaps in beliefs about public health measures -- was surprising.

However, while the logic of partisan media consumption driving partisan differences in belief about Covid-19 public health measures seems sound, empirically proving the correlation at a fine-grained level is much more difficult. In general, quantitative study of wide-reaching media effects on public attitudes is a challenging endeavor, though some have taken the time to do so. Early in the Covid-19 pandemic, Bursztyn et al. correlated media consumption data with county-level Covid-19 case rates, arguing that consumption of right-leaning opinion shows like \emph{Tucker Carlson Tonight} or \emph{Hannity} correlated with higher case rates \cite{Bursztyn2020MisinformationPandemic}. Commenting on an earlier case of partisan differences in belief, Benkler, Faris and Roberts demonstrated that certain major beliefs affecting the 2016 U.S. presidential election correlated with the \emph{number of articles} published by partisan media sources and their hyperlinking patterns, but only offer qualitative detail of what was contained \emph{within} the stories \cite{Benkler2018NetworkPolitics}. Other studies have used natural language processing techniques such as topic modeling \cite{Scoville2022MaskReader,He2021WhyPandemic,Taylor2021NegativeReactance,Al-Ramahi2021PublicData,Pascual-Ferra2021ToxicityPandemic,Krawczyk2021QuantifyingResource,Lee2021StructuralData,Shin2022Mask-WearingElites} or simple keyword searches \cite{Brady2019AnLeaders.,Koevering2022ExportingDiscussed} to describe media messaging or social media discourse related to Covid-19, but even these studies lose much of the detail contained within media stories as they are summarized into topics or keyword bundle matches.

To address this gap, this study demonstrates a novel method to structure and analyze media text to show the influence of paragraph-level media messaging on public opinion. We do so by creating a dataset of major U.S. media messaging surrounding Covid-19 mask-wearing -- annotated along 14 dimensions of mask-wearing attitudes taken from Howard 2020's validated Face Mask Perceptions Scale (FMPS) \cite{Howard2020UnderstandingScale} -- and using the resulting data to replicate polling data presented by Gallup's Bird \& Ritter \cite{Bird2020IsPractices} demonstrating the formation of a partisan gap in mask-wearing attitudes over time in the early pandemic.

One aspect of this contribution is the use of a validated psychological questionnaire to create a code book that annotators could use to identify mask-wearing attitudes in media data. Howard's scale provides an excellent set of validated attitudes concerning Covid-19 mask-wearing across multiple dimensions, which made the task of deciding on labels for the data trivial, as they were drawn from a rigorous scientific source. The resulting dataset, titled \emph{FMPS Attitudes in U.S. Covid-19 News}, consists of 2,562 high-quality annotations with high inner-annotator agreement (181 human-labeled news paragraphs across 119 articles, with 14 annotations for each paragraph), spanning the dates of April 6 to June 8, 2020, on news stories from five major U.S. news sources and three prime time opinion shows than span the U.S. ideological and partisan spectrum. 

This method and dataset allowed us to subsequently conduct a quantitative analysis of pro- and anti-mask sentiment by taking aggregate scores for each annotated article across the 14 FMPS dimensions we use from Howard 2020 \cite{Howard2020UnderstandingScale}. We use these sentiment-like ratings to identify time spans where pro- and anti-mask scores differed significantly between different media diets, and then conduct a deeper analysis of the annotated messages in these time spans. For the cases we identify and highlight, we find that rhetoric tends to vary as one may expect between Democrat and Republican media diets: the former includes messages advocating for mask-wearing while the latter casts doubts on masks as a public health measure. But notably, we find some exceptions: \emph{Hannity}, early in the dataset, contains several pro-mask messages; and outlets in the Democrat media diet, though yielding pro-mask scores in the aggregate of their 14 FMPS dimensional scores, are shown to contain messages claiming, for example, that face masks are uncomfortable or may not be effective -- but not as many as from Republican diet outlets. From our qualitative analysis of media messages, we find that pro-mask messages typically highlight the efficacy of masks. In contrast, we find that anti-mask messages cast doubt on mask efficacy, question the trustworthiness of public health officials, frame the issue as one of partisan ``us'' versus ``them,'' and critique mask mandates as infringement on individual liberties. 

We also use this dataset to demonstrate a case where opinion in aggregate can, in fact, be shown to have changed in accordance with attitudes expressed in specific mainstream news stories at the granularity of paragraphs. Using a simple opinion change model updating aggregate opinion values for groups with different media diets (Democrat media consumers, moderate, and Republican) according to mask-wearing attitudes found in media articles along the dimensions of the FMPS, we generate simulated time series opinion data that qualitatively resembles that of Bird \& Ritter's Gallup poll of mask-wearing support taken between the same days \cite{Bird2020IsPractices}. Our results show that different media diets' messaging around mask-wearing was enough, under this simple model, to create and maintain a partisan divide in support for masking.

Beyond our one simple model validation, the \emph{FMPS Attitudes in U.S. Covid-19 News} dataset can serve as a sandbox for other opinion change models and studies to validate their dynamics against real, empirical inputs and outputs. In addition to making our labeled news dataset open to the public and available for further study (with a snapshot at \cite{Rabb2023FaceMaskData}; and a continually updated dataset at https://github.com/ricknabb/media-ideology-coding). Moreover, this method for data structuring can be used by other media researchers seeking to understand the interplay between media messaging and opinion formation at a fine-grained level. We hope that other researchers will be able to replicate and refine this annotation method and use it to describe other attitudes relevant to belief in media information (e.g., trust and source credibility \cite{Pennycook2021TheNews,Ecker2022TheCorrection,Schwarz2021WhenMisinformation} or group identity signaling \cite{VanBavel2018TheBelief,Oyserman2021YourAccept}). There are several ways that this work could be extended, so we conclude by discussing these directions and others that can work towards more detailed, systematic study of media attitudes and their influence on important public beliefs.

\section{Background}

\subsection{Quantitative study of media effect on opinion}

Study of the influence of media on opinion is an old and rich area, with seminal works appearing in the early 20th century from the likes of media theorists and early psychologists like Lippmann \cite{Lippmann1922PublicOpinion}, Bernays \cite{Bernays1928Propaganda}, and others, but has remained a difficult inquiry to conduct with detailed data and models. Quantitative study of the effect of media on opinion formation is a goal of many researchers and practitioners, and underlies diverse applications such as marketing, innovation prediction \cite{Kiesling2012Agent-basedReview}, and political polling \cite{Converse2017Survey1890-1960}, to name a few. 

Recently, in the area of studying political beliefs, particular interest has been taken in understanding the influence of media on polarized attitudes, such as partisan beliefs in the U.S. during the 2016 presidential election \cite{Kuo2021CriticalPolitics}, and during the Covid-19 pandemic \cite{Murphy2023What2016-2022}. Yet to study the direct effect of media messaging on attitudes is perhaps even more difficult than it once was, as social media and the Internet have allowed individuals to consume a wider and more diverse array of media sources than before these technological changes \cite{Karlsen2020Do19972016}.

Political polling and research organizations, like Pew and Gallup in the U.S., conduct opinion polls to gauge political attitudes of the population and how they change over time. Recently, these organizations have published reports demonstrating partisan polarization in topics such as racial discrimination \cite{Hurst2023AmericansExist}, climate change \cite{Pasquini2023AmericansChange}, and Covid-19 \cite{Gramlich2020202020,Newport2020TheCoronavirus}. A Gallup study by Bird \& Ritter \cite{Bird2020IsPractices} was able to show significant partisan differences in opinion polls about Covid-19 health behaviors, and argue they are driven by media consumption by comparing which sources they mainly turn to for news. These studies provide useful data showing population-level beliefs over time.

Some studies perform more detailed analysis of media stories themselves, arguing about their influence on political attitudes. Bursztyn et al. \cite{Bursztyn2020MisinformationPandemic} demonstrated relationships between county-level differences in Covid-19 cases and viewership of Fox opinion shows, quantifying the Covid-related messaging from the shows using keyword searches and crowd-sourced coding. In the early pandemic, Kouzy et al. \cite{Kouzy2020CoronavirusTwitter} also hand-coded topics in tweets, but this method proves difficult to scale to large data. Benkler, Faris, and Roberts, in \emph{Network Propaganda}, performed a landmark analysis of social and traditional media consumption, across large media networks, and how they led to belief in disinformation surrounding the 2016 U.S. presidential election \cite{Benkler2018NetworkPolitics}. While their studies captured the scale of the media ecosystem, their quantification relied on keywords, reducing the complexity with which they could study individual stories' influence on attitudes.

Other researchers have turned to opinion diffusion modeling as a method to reason about polarization and change in public opinion. Diffusion models capture the dynamic exchange of opinion throughout simulated social networks where nodes hold beliefs and are influenced by their connections. Many opinion diffusion models study the formation of polarized populations \cite{Dandekar2013BiasedPolarization,DellaPosta2015WhyLattes,Goldberg2018BeyondVariation,Sikder2020ANetworks,Rabb2023CognitivePolarization}, with some validating their final opinion distributions against empirical opinions from polling \cite{Duggins2017APolitics}. However, while several innovation diffusion models empirically validate their results \cite{Zhang2019EmpiricallyReview}, it is more rare for those modeling political opinion polarization to do so, as the data necessary to do so can be difficult to obtain.

\subsection{Crowd-sourced data annotation}

Gathering data on complicated interpretations of text, images, or video is a difficult task. There can be competing interpretations of information, with some potentially more accurate than others. Moreover, labeling data is a costly endeavor both in terms of time and money, and can be prohibitively challenging for one or a few annotators.

One solution to these challenges is to employ the wisdom of a crowd in annotating large datasets \cite{Pustejovsky2012NaturalApplications}. Even if annotators have somewhat different interpretations of the meaning of data, with enough individuals annotating the same sample, a signal starts to emerge from the noise. Additionally, several techniques have been developed to gauge the quality of annotations. For example, an piece of data annotated by ten individuals can be said to have a high-quality label if the annotators strongly agreed with each other on the label -- what is called a high \emph{inter-rater reliability}. Typically, this agreement is measured with Fleiss' Kappa \cite{Landis1977AnObservers} (described in Eq. \ref{eq:fleiss-kappa}), which gives a degree of agreement among raters while also making sure the agreement is not from chance.

\begin{align}
    \kappa = \frac{p_o - p_e}{1 - p_e}
    \label{eq:fleiss-kappa}
\end{align}

Crowd-sourced annotation methods have been used for a variety of applications, including for labeling media and social media data. Studies have used the method to label rumors on social media \cite{Zubiaga2015CrowdsourcingMedia}, biased language in the news \cite{Lim2020AnnotatingCrowdsourcing}, and sentiment in online articles \cite{Brew2010UsingMedia}. So long as annotations are evaluated for their quality, the method can be a powerful tool to gather data that may otherwise be too difficult for small teams of researchers.

\section{Methods}

\subsection{Data collection}

To gather mainstream U.S. news data, we utilized the MediaCloud platform to download news items from five media sources spanning both Democratic and Republican viewership, and additionally scraped transcripts for three sources. We collected written news stories from these outlets (for opinion shows, we pulled transcripts posted online) that were published between April 6, 2020 and June 8, 2020. The full set of stories is available for download at \cite{Rabb2023FaceMaskData}. 

MediaCloud is a platform that collects web-based news articles over time, and has amassed a vast collection spanning a large number of media producers since its launch. It allows users to query articles using filters, and returns stories based on boolean search queries that match text taken from the HTML body of the page. The query that we used to find stories was: \textbf{(`covid' or `coronavirus' or `covid-19' or `virus') and `mask'}. Data for the \emph{New York Times}, \emph{Fox News}, \emph{Breitbart}, \emph{Tucker Carlson Tonight}, \emph{The Ingraham Angle}, and \emph{Hannity} was accessed in June, 2021; while data for \emph{Daily Kos} and \emph{Vox} was accessed in July, 2023.

These outlets and the time span we searched within were chosen to match a Gallup study conducted to gauge behaviors and attitudes relating to Covid-19 public health measures from the same time range \cite{Bird2020IsPractices}. This study, whose methodology can be found in the full article, polled individuals as groups based on different media diets, where the Democrat media diet included left-leaning sources (e.g., \emph{New York Times}, \emph{Vox}, or \emph{Daily Kos}); moderate media diets included a mix of left- and right-leaning sources (e.g., \emph{New York Times} and \emph{Fox News}); and Republican media diets included right-leaning sources (e.g., \emph{Breitbart}, \emph{Fox News}, or \emph{Tucker Carlson Tonight}). To mirror this methodology, we assigned media outlets into distinct Democrat, Moderate, and Republican media diets, listed in Table \ref{tab:news-diets}.

\begin{table}[h!]
    \centering
    \begin{tabular}{c|c}
        Democrat diet  &  Daily Kos, Vox, New York Times\\
        \hline
        Moderate diet  &  New York Times, Fox News\\
        \hline
        Republican diet & Fox News, Breitbart, Tucker Carlson Tonight,\\
        & The Ingraham Angle, Hannity\\
    \end{tabular}
    \caption{News diets by partisan affiliation}
    \label{tab:news-diets}
\end{table}

\begin{figure*}[!h]
    \centering
    \begin{subfigure}{0.45\textwidth}
        \centering
        \includegraphics[width=\linewidth]{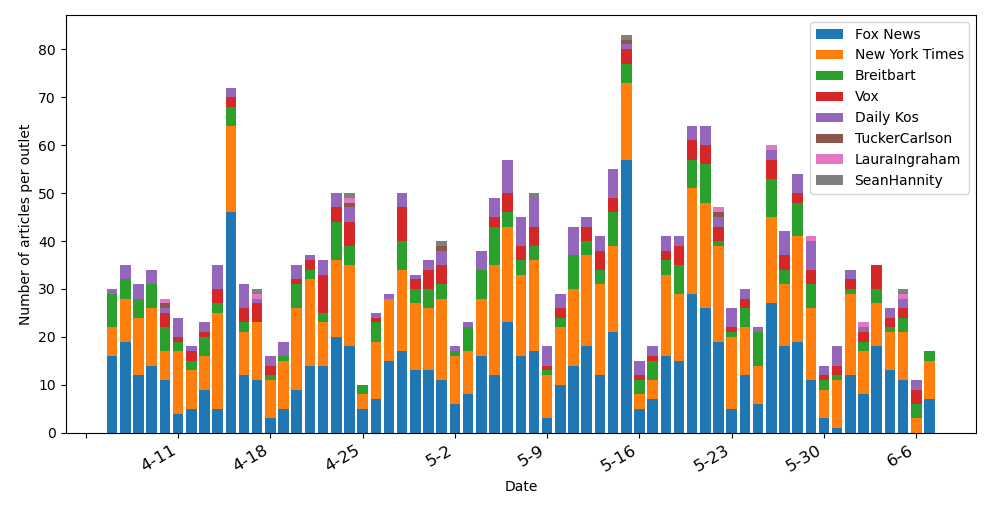}
        \caption{The number of articles gathered over time, in a stacked plot labeled by media outlet.}
        \label{fig:articles-over-time-outlet}
    \end{subfigure}
    \begin{subfigure}{0.45\textwidth}
       \centering
       \includegraphics[width=\linewidth]{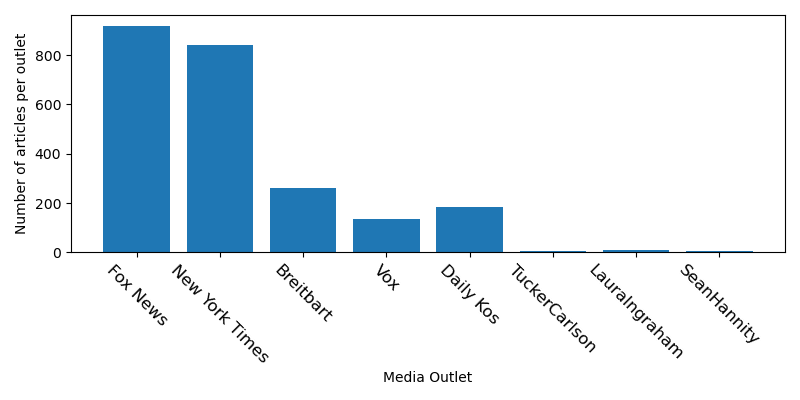}
       \caption{The number of articles from each media outlet.}
       \label{fig:articles-per-outlet}
   \end{subfigure}
\end{figure*}

In total, we gathered 2,361 news articles, yielding 8,473 total paragraphs. The number of articles over time and labeled by media outlet is displayed in Fig. \ref{fig:articles-over-time-outlet}, and the number of articles gathered per media outlet is displayed in Fig. \ref{fig:articles-per-outlet}.

\subsection{Annotation}

To reach our goal of creating an annotated dataset of mask-wearing statements across these mainstream U.S. media outlets, we employed a crowd-sourced annotation method to label the text gathered from the above process. But as opposed to many crowd-sourced methods which use online platforms like MTurk to find annotators, we gathered participants from our university campus for in-person annotation sessions, which guarantees a higher quality of labels.

Annotators were presented with single paragraph-sized pieces of text from full articles and asked to identify whether they were expressing certain Covid mask-wearing attitudes. We chose to have annotators label paragraphs so that in future applications, automated systems using these labels to subsequently label larger datasets would be able to try to find similar signals in blocks of text with guaranteed (albeit variable) length. Moreover, by annotating at the paragraph scope, articles could be analyzed while incorporating, for example, variable attention levels based on where paragraphs are situated in the whole article (earlier paragraphs may be paid more attention as opposed to the final paragraphs). We discuss in more detail the implications of paragraph-level annotation in Section \ref{sec:limitations} Limitations and Future Work.

To gauge a comprehensive set of attitudes, we utilized a validated psychological questionnaire on mask-wearing attitudes from Howard 2020, the Face Mask Perception Scale \cite{Howard2020UnderstandingScale}, which identifies 32 attitudes that are relevant to mask-wearing belief. We consolidated items from the questionnaire that we considered highly related (e.g., ``face masks disrupt my breathing'' and ``it is difficult to breathe when wearing a face mask''), which resulted in 14 attitudes that we presented to annotators. The full list of questions presented to annotators, as well as their overarching categorical classification, is displayed in Table \ref{tab:mask-coding-questions}.

\begin{table*}[!h]
    \centering
    \begin{tabular}{l|c}
        Coding Prompt & Labels\\
        \hline
        \hline
        \multicolumn{2}{l}{Does the piece of text presented convey the idea that...}\\
        \multicolumn{2}{l}{\textbf{Comfort}}\\
        \hline
        \hline
        ...it is difficult to breathe while & 0 = it is difficult, 1 = it is not difficult,\\
        wearing a face mask? & 2 = does not mention \\
        \hline
        ...face masks get too hot? & 0 = they get too hot, 1 = they do not get too hot,\\
        & 2 = does not mention\\
        \hline
        \multicolumn{2}{l}{\textbf{Efficacy}}\\
        \hline
        \hline
        ...face masks provide health benefits? & 0 = they provide few health benefits,\\
        & 1 = they provide health benefits, 2 = does not mention \\
        \hline
        ...face masks are effective? & 0 = they are ineffective, 1 = they are effective,\\
        & 2 = does not mention\\
        \hline
        \multicolumn{2}{l}{\textbf{Access}}\\
        \hline
        \hline
        ...it is difficult to get a face mask? & 0 = it is difficult, 1 = it is easy,\\
        & 2 = does not mention \\
        \hline
        ...face masks are too expensive? & 0 = they are too expensive, 1 = they are affordable,\\
        & 2 = does not mention\\
        \hline
        \multicolumn{2}{l}{\textbf{Compensation}}\\
        \hline
        \hline
        ...you can simply stay away from & 0 = you can stay away from people,\\
        people when you go out? & 1 = you cannot stay away from people, 2 = does not mention \\
        \hline
        \multicolumn{2}{l}{\textbf{Convenience}}\\
        \hline
        \hline
        ...people do not like & 0 = they do not like remembering,\\
        remembering to wear a face mask? & 1 = they do not mind remembering, 2 = does not mention\\
        \hline
        ...wearing a face mask is too much & 0 = it is too much of a hassle,\\
        of a hassle? & 1 = it is not too much of a hassle, 2 = does not mention\\
        \hline
        \multicolumn{2}{l}{\textbf{Appearance}}\\
        \hline
        \hline
        ...face masks look ugly or weird? & 0 = they look ugly or weird,\\
        & 1 = they do not look ugly or weird, 2 = does not mention\\
        \hline
        \multicolumn{2}{l}{\textbf{Attention}}\\
        \hline
        \hline
        ...face masks make people & 0 = they make people seem untrustworthy,\\
        seem untrustworthy? & 1 = they do not make people seem untrustworthy, 2 = does not mention\\
        \hline
        ...face masks make others & 0 = they make others uncomfortable,\\
        uncomfortable? & 1 = they do not make others uncomfortable, 2 = does not mention\\
        \hline
        \multicolumn{2}{l}{\textbf{Independence}}\\
        \hline
        \hline
        ...people do not like being & 0 = they do not like being forced to do something,\\
        forced to do something? & 1 = they are willing to be forced to do things, 2 = does not mention\\
        \hline
        ...people want to prove a point & 0 = they want to prove a point against authority,\\
        against authority? & 1 = they are willing to follow authority, 2 = does not mention\\
        \hline
    \end{tabular}
    \caption{A table containing the prompts given to crowd-sourced coders for a subset of mask-wearing belief factors, and its corresponding answers coded into label values.}
    \label{tab:mask-coding-questions}
\end{table*}

Annotators were presented with paragraphs and questions side-by-side in a custom React.js web application used for recording annotator responses. Each paragraph was annotated according to the 14 attitudes and annotators additionally assigned a confidence score to each response on a 7-point Likert scale ranging from ``Not confident'' (1) to ``Unsure'' (4) to ``Very confident'' (7). For each attitude, the question asking whether the attitude was present was able to be answered with three options for annotation: the attitude is present, the opposite of the attitude is present, or neither is present.

\begin{figure}[!h]
    \centering
    \includegraphics[width=.9\linewidth]{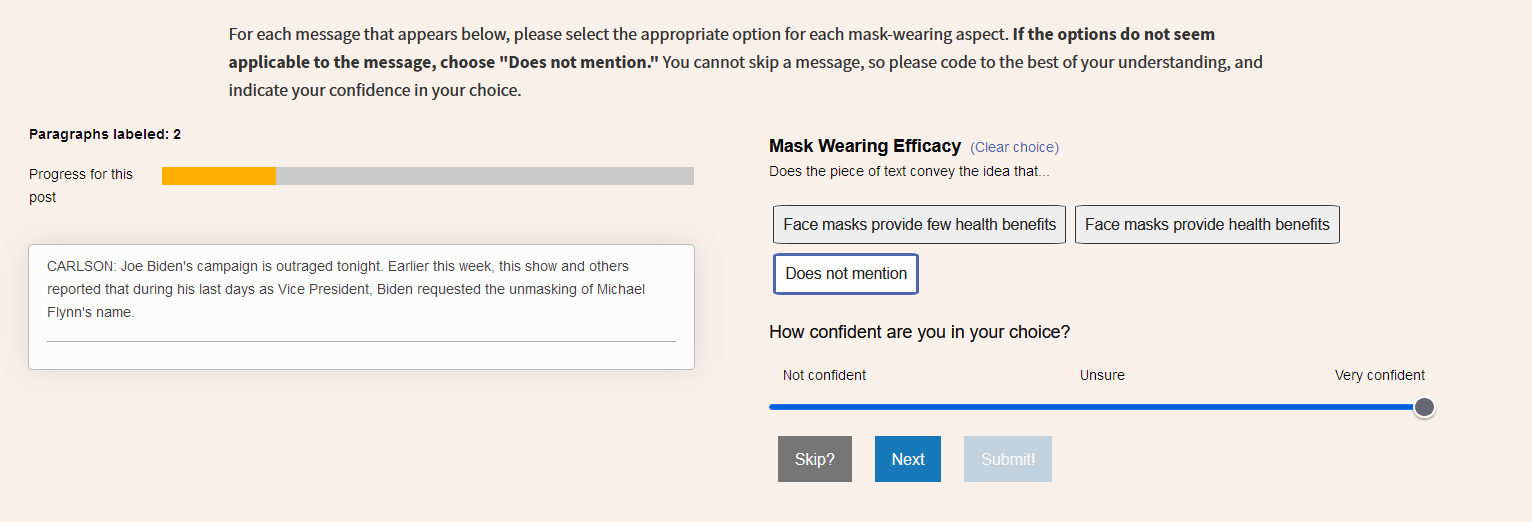}
    \caption{A screenshot of the web tool that participants used for annotating snippets of news articles.}
    \label{fig:coding-tool}
\end{figure}

Before proceeding to annotate data from our dataset, participants completed a training task which asked them to annotate five example paragraphs -- three of which were fabricated, and two of which were from the real dataset. Training questions can be found in the Supplemental Materials. A researcher was present during the training task to allow participants to ask clarifying questions about using the tool or the meaning of the prompts, while attempting to not influence the participant's annotations. After completing the training task, participants requested permission from the present researcher to proceed to annotating data from the real dataset.

Participant identifiers were not collected during this study. Each annotator was assigned a unique ID that was tracked in their browser local storage to maintain their session if the page refreshed or they lost internet connection momentarily. The entirety of the dataset was collected over several rounds of in-person annotation, resulting in 22 unique annotator IDs in the dataset (meaning, at most, 22 individuals labeled the data, but it may be fewer as some annotators likely cleared their browser storage between sessions).

\section{Results}

\subsection{FMPS Attitudes in U.S. Covid-19 News dataset}

\begin{figure*}[!h]
    \centering
    \begin{subfigure}{0.4\textwidth}
        \centering
        \includegraphics[width=\linewidth]{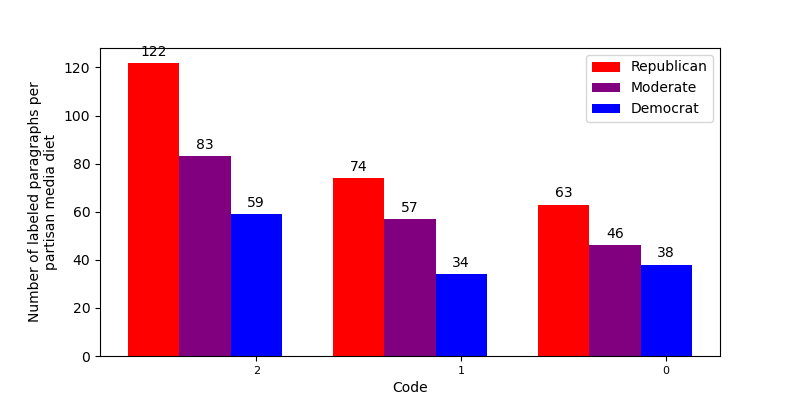}
        \caption{High-quality labels for the dataset.}
    \end{subfigure}
    \begin{subfigure}{0.55\textwidth}
        \centering
        \includegraphics[width=\linewidth]{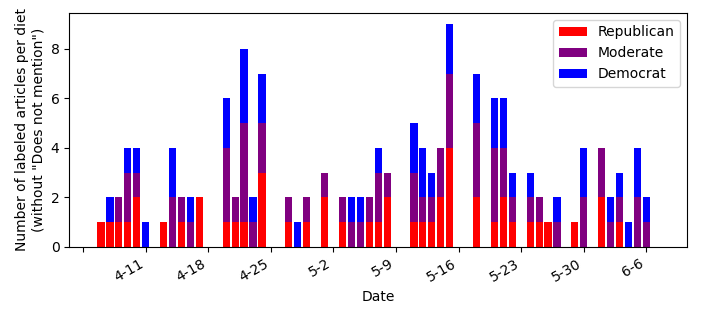}
        \caption{Labels over time for different partisan media diets.}
    \end{subfigure}
    \caption{Distributions of labels for both the set of all codes, and high-quality codes, grouped by partisan media diet.}
    \label{fig:label-partisan-distribution}
\end{figure*}

Our final dataset, which we call \emph{FMPS Attitudes in U.S. Covid-19 News}, resulted in 7,559 total annotations across 297 of the 8,473 total paragraphs, and 202 of the 2,361 articles we collected. The number of annotations by article date and by media diet are shown in Fig. \ref{fig:label-partisan-distribution}. Additionally, several paragraphs were labeled by multiple participants. All data and annotations are available, and publicly updated with new results, on GitHub at https://github.com/ricknabb/media-ideology-coding. A snapshot of this dataset at the point used for the following analysis is available at \cite{Rabb2023FaceMaskData}.

\begin{figure*}[!h]
    \centering
    \includegraphics[width=\linewidth]{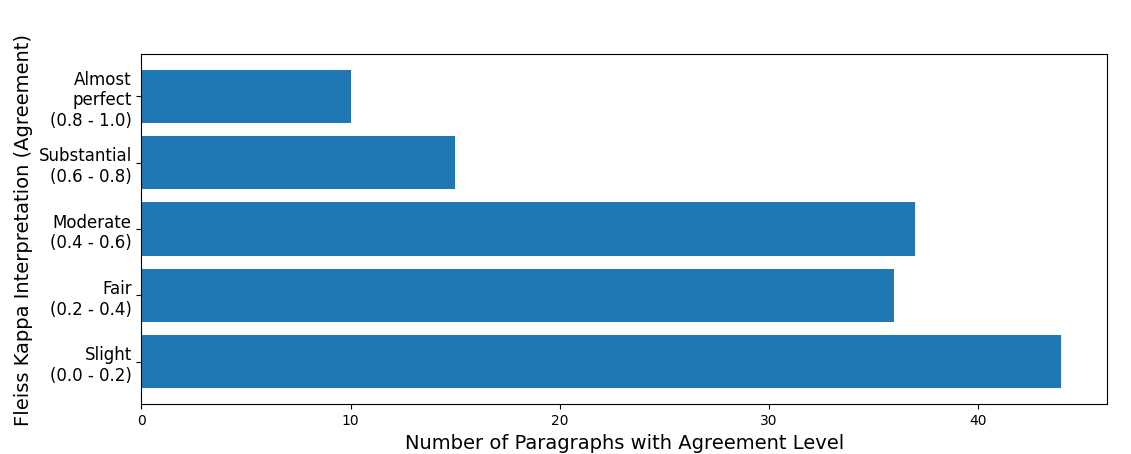}
    \caption{Frequency of Fleiss' Kappa scores for paragraphs annotated by more than two annotators, broken into categories according to Landis \& Koch \cite{Landis1977AnObservers}.}
    \label{fig:kappa-distribution}
\end{figure*}

To gauge the quality of annotations assigned to paragraphs and remove those which were low-quality, we employed two filters. First, for paragraphs that were only labeled by one participant, we removed annotations which were marked with a confidence score below 5. Then, for paragraphs which were annotated by multiple annotators, we calculated Fleiss' kappa and only kept annotations which yielded an agreement score greater than or equal to 0.4 (which is considered by Landis \& Koch \cite{Landis1977AnObservers} to be ``moderate'' agreement). A distribution of all inter-rater agreement scores is shown in Fig. \ref{fig:kappa-distribution}.

As evidenced by the distribution of kappa scores, just over half the paragraphs annotated by multiple annotators have low scores (fair, 0.2 - 0.4; or slight, 0.0 - 0.2). There is a question as to whether mapping self-reported psychological attitudes to media speech is a valid application of one type of content to another. We note that this is a different technique than, for example, content analysis which uses topic models to find attitudes \emph{a posteriori} (our use of a code book assumes \emph{a priori} that the attitudes may exist in media text). Based on the number of high scores (greater than 0.4) versus low scores, we conclude that there may be difficulties applying Howard 2020's \cite{Howard2020UnderstandingScale} FMPS to all media text. This may be due either to the possibility that self-reported psychological attitudes may not straightforwardly map onto media speech, or that there may be varying degrees of ambiguity in media speech that prevents high inter-rater agreement \emph{regardless} of the code book used. We present a sample of low-score and high-score paragraphs in Table \ref{tab:high-low-kappa-paragraphs} with their associated kappa score.

\begin{table*}[!h]
    \centering
    \begin{tabular}{p{0.1\linewidth} | p{0.9\linewidth}}
        \textbf{Fleiss Kappa Score} & \textbf{Paragraph Text}\\
        \hline
        \multicolumn{2}{l}{High-Scoring Paragraphs}\\
        \hline
        1.0 & On Wednesday, President Donald Trump said during the White House coronavirus daily briefing that the nation's Strategic National Stockpile of personal protective equipment (PPE), such as N95 masks, had been nearly depleted when he assumed office.\\
        \hline
        0.868 & Researchers have found that wearing surgical masks can significantly reduce the rate of airborne COVID-19 transmission, according to a study released on Sunday.\\
        \hline
        0.65 & And in Houston, Governor, if you can believe this, they are threatening to punish people for not wearing a mask. Meanwhile, on Washington state, they are going to let a serial killer, they are one vote away from letting a serial killer out because of COVID-19. \\
        \hline
        \multicolumn{2}{l}{Low-Scoring Paragraphs}\\
        \hline
        0.391 & President Trump himself has taken heat for not wearing a mask in public, even as one of his valets and a staff member for Vice President Mike Pence have tested positive for the coronavirus. Birx said Trump wore a mask for a period of time at a Ford plant last week.\\
        \hline
        0.251 & Breaking today, the CDC is now recommending that all Americans wear face masks, that's new today, or other facial coverings when traveling outside the home. And it's all comes as the president utilizes the Defense Production Act to ban certain exports of critical medical supplies. He did this because 3M -- they were selling masks made -- American-made masks, N95 masks, to other countries. Hello? \\
        \hline
        0.063 & Another NYPD officer stationed in Brooklyn was also suspended without pay after pulling down a protester's mask and pepper spraying him in the face on May 30. The incident was captured on video and shared to Twitter, where it's received more than 3 million views so far. \\
    \end{tabular}
    \caption{A table of high-scoring and low-scoring paragraphs that were coded by multiple annotators, according to their Fleiss kappa scores broken into categories according to Landis \& Koch \cite{Landis1977AnObservers}. Low-scoring paragraphs received a score less than 0.4, and high-scoring greater than or equal to 0.4.}
    \label{tab:high-low-kappa-paragraphs}
\end{table*}

It is unclear from the small sample of paragraphs and agreement scores if lower agreement scores correlate with more ambiguous or confusing text, or if there is simply a general difficulty in applying self-reported attitudes to media text. We leave a more rigorous and comprehensive evaluation of the justifiability of applying FMPS attitudes to media text to future work. For the purposes of this study, we decide to only use data from paragraphs coded by several annotators that yielded high agreement score (greater than or equal to 0.4). To resolve paragraphs coded by multiple annotators that passed this agreement test, we took a majority vote for each attribute that was coded, and in the case of a tie, chose the label with the highest confidence score. If confidences are equal, then a random code is chosen. The final number of high-quality annotations was 2,562 across 181 paragraphs and 119 articles.

In addition, for each paragraph and article, we can calculate a pro- or anti-mask belief score according to the annotations given for each of the mask-wearing attributes. A paragraph would be rated as follows:

\begin{align}
    \pi(L) &= \frac{0.5 + positive(L)}{1 + positive(L) + negative(L)},
    \label{eq:zero-one-article-rating}
\end{align}

\noindent where $L$ is the set of labels on the paragraph, and $positive(L)$ or $negative(L)$ are how many annotations of 1 or 0 were given, respectively. An article is coded as simply the average of all paragraph scores for that article.

\begin{figure*}[!h]
    \centering
    \includegraphics[width=\linewidth]{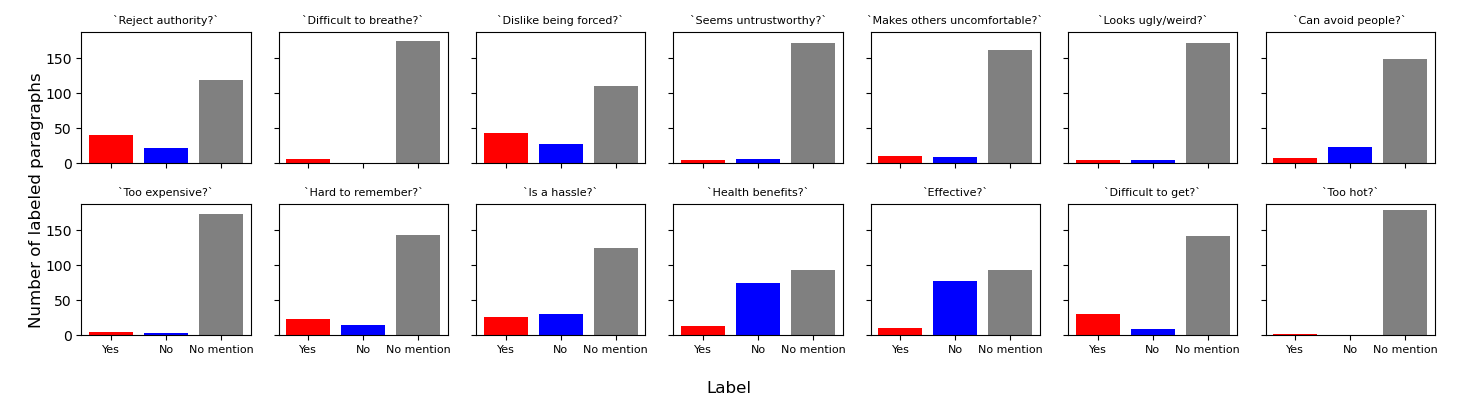}
    \caption{Distributions of labels by all mask-wearing attributes listed in Table \ref{tab:mask-coding-questions}. Pro-mask labels are colored in blue, anti-mask labels in red, and ``does not mention'' in gray.}
    \label{fig:labels-by-attribute}
\end{figure*}

\begin{figure}[!h]
    \centering
    \begin{subfigure}{1.0\textwidth}
        \includegraphics[width=\linewidth]{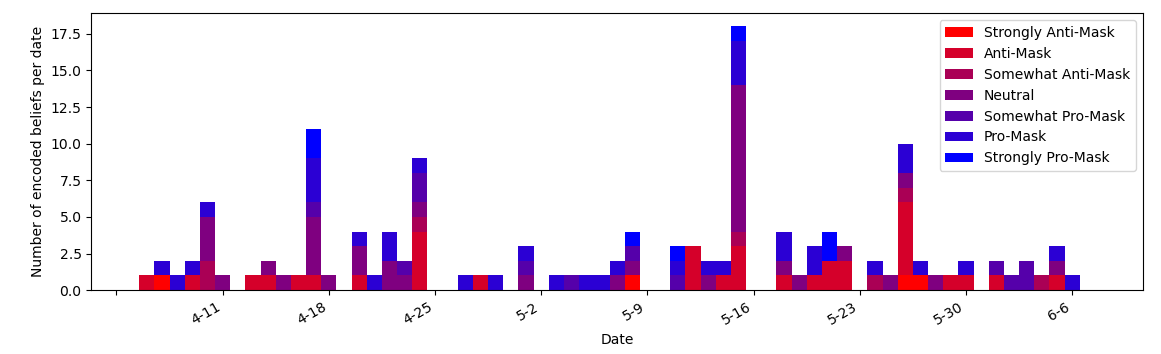}
        \caption{Stated belief in mask-wearing, from ``Strongly pro-mask'' to ``Strongly anti-mask,'' as number of paragraphs with those sentiments over time.}
        \label{fig:article-belief-over-time}
    \end{subfigure}
    \begin{subfigure}{1.0\textwidth}
        \includegraphics[width=\linewidth]{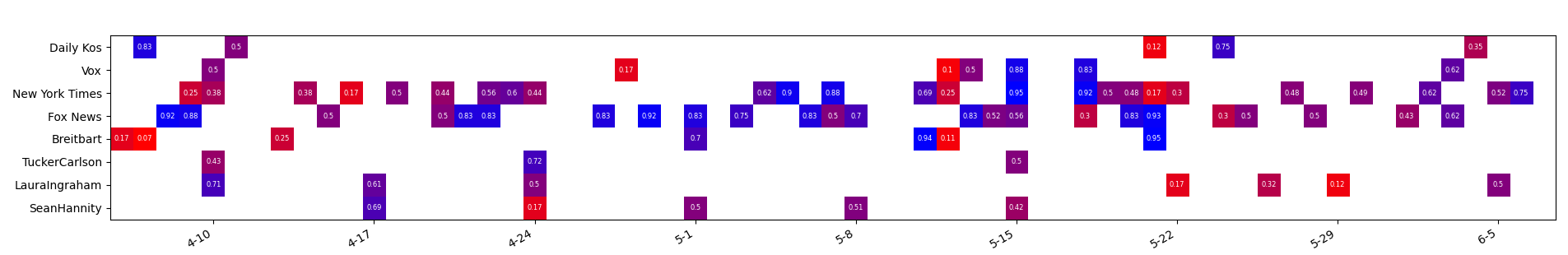}
        \caption{Average belief scores from 0 to 1 (where 0 is strongly anti-mask and 1 is strongly pro-mask) per media outlet for each day in the dataset.}
        \label{fig:media-article-belief-over-time}
    \end{subfigure}
\end{figure}

Using this measure, we can characterize the frequency of pro- and anti-mask statements over time, as displayed in Fig. \ref{fig:article-belief-over-time}. We can also visualize the mean belief value for each day, per media outlet, for days that they published articles about mask-wearing. Those results can be viewed in Fig. \ref{fig:media-article-belief-over-time}.

\subsection{Simulating opinion change for partisan media consumers}

To conduct a simple test on the dataset, we ran a model of opinion change to compare its output to polling results from Bird and Ritter \cite{Bird2020IsPractices}. Their study reported divergent mask-wearing attitudes for partisan media consumers across the first two months of the Covid-19 pandemic. But their analysis did not comment on the content that viewers consumed, only on the effect. To note, timeseries opinion data that we compare against from \cite{Bird2020IsPractices} was not available for redistribution, so we estimated data values from a visualization of their results. More details of this data estimation process can be found in Supplemental Materials.

Our simple model, which we will call the \emph{Single Point Model}, was initialized with three initial opinion values taken from \cite{Bird2020IsPractices} -- a Democrat diet opinion, moderate diet, and Republican diet (with sources listed above in Table \ref{tab:news-diets}) -- as percentages (in the Bird \& Ritter study, the percent of media consumers who responded as having worn their face mask in the past seven days). For each day from April 6 to June 8, 2020, labeled articles were given a rating measuring their mask-wearing attitudes as the mean value of each of its labeled paragraph ratings. Based on an average belief score $\overline{\pi(L)}$ for an article $a \in A$, the opinion score would be changed according to a simple decision function:

\begin{equation}
    \Delta_O(a) =
        \begin{cases}
        +1, & \overline{\pi(L)} > 0.5,\\
        +0, & \overline{\pi(L)} = 0.5,\\
        -1, & \overline{\pi(L)} < 0.5
        \end{cases}
    \label{eq:opinion-score}
\end{equation}

At each date in the dataset, any annotated article would contribute to changing opinion values. Based on the different media diets, each group would update their value if that media source was part of their diet. For any diet $D$, its opinion value $O$ at time $t+1$ would update based on the sum of article scores written by source $d \in D$ at time $t$:

\begin{equation}
    O(D, t+1) = O(D,t) + \sum\limits_{a \in A_{d, t}}{\Delta_O(a)}
    \label{eq:update-rule}
\end{equation}

We detail the Single Point Model, including its initial values and update rules, in Table \ref{tab:single-point-model}.

\begin{table}[h!]
    \centering
    \begin{tabular}{p{0.4\linewidth} | p{0.6\linewidth}}
        \textbf{Model Parameter} & \textbf{Value}\\
        \hline
        Republican media diet & Fox News, Breitbart, Tucker Carlson Tonight, The Ingraham Angle, Hannity\\
        \hline
        Moderate media diet & New York Times, Fox News\\
        \hline
        Democrat media diet & Daily Kos, Vox, New York Times\\
        \hline
        Initial Republican opinion & 43.12\%\\
        \hline
        Initial Moderate opinion & 48.58\%\\
        \hline
        Initial Democrat opinion & 65.34\%\\
        \hline
        Update rule & $O(D, t+1) = O(D,t) + \sum\limits_{a \in A_{i, t}}{\Delta_O(a)}$
    \end{tabular}
    \caption{The Single Point Model used to simulate changes in opinion given our annotated dataset.}
    \label{tab:single-point-model}
\end{table}

\begin{figure*}[!h]
    \centering
    \begin{subfigure}{0.45\textwidth}
        \centering
        \includegraphics[width=\linewidth]{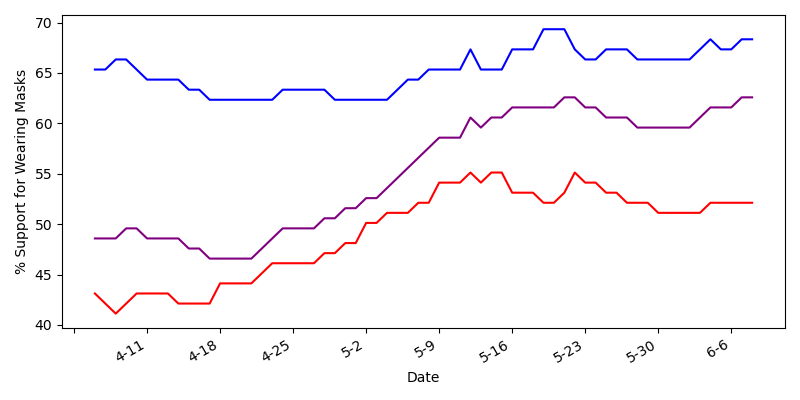}
        \caption{Simulation on high-quality codes with paragraph scores averaged across an article}
        \label{fig:opinion-sim-hq}
    \end{subfigure}
    \begin{subfigure}{0.45\textwidth}
        \centering
        \includegraphics[width=\linewidth]{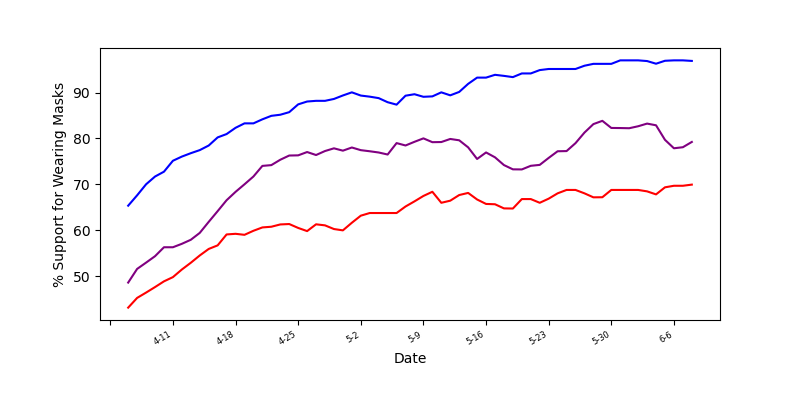}
        \caption{Empirical polling data from Bird \& Ritter's Gallup study \cite{Bird2020IsPractices}}
        \label{fig:opinion-sim-gallup}
    \end{subfigure}
    \begin{subfigure}{1.0\textwidth}
        \includegraphics[width=\linewidth]{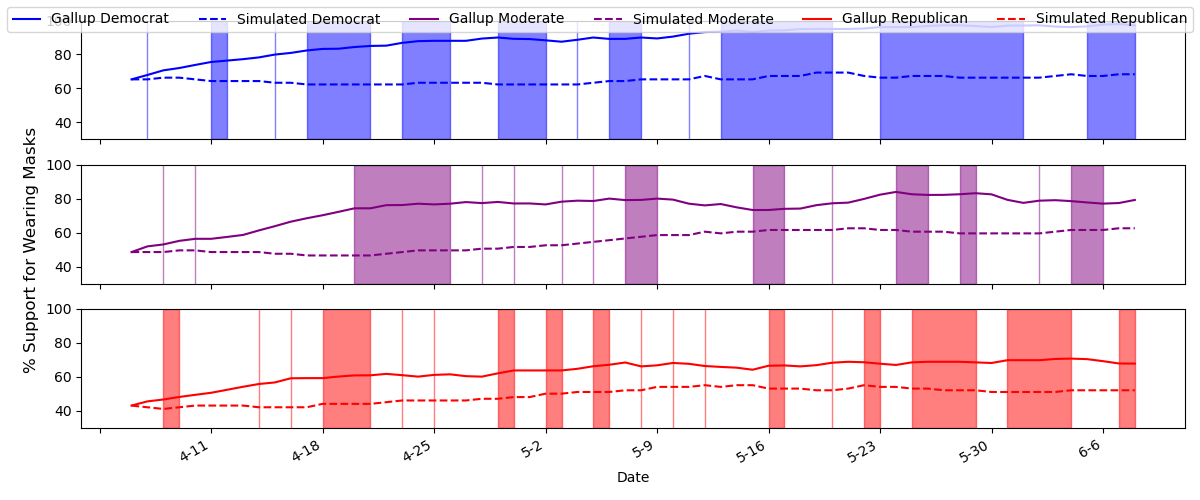}
        \caption{Areas where the rate of change of the simulated data and empirical data match within a threshold (described in Eq. \ref{eq:timeseries-similarity}.).}
        \label{fig:gallup-sim-comparison}
    \end{subfigure}
    \caption{Results of simple opinion change simulations on the set of all codes and of high-quality codes, compared with data from Gallup polling about mask-wearing opinion over time.}
    \label{fig:opinion-simulation}
\end{figure*}

After simulating opinion change for our high-quality annotated articles, the results are displayed in Fig. \ref{fig:opinion-simulation}.

We note that there is a decent qualitative similarity to the patterns of our simulation and those found by the Gallup poll. Fig. \ref{fig:gallup-sim-comparison} plots time ranges where the simulated data and empirical data have similar rates of change, as described in Eq. \ref{eq:timeseries-similarity}.

\begin{equation}
    highlight(f_e, f_s, t) = \Delta(f_e,t) * \Delta(f_s,t) > 0\ \lor\ |\Delta(f_e,t) - \Delta(f_s,t)| \leq 1,
    \label{eq:timeseries-similarity}
\end{equation}

\noindent where $\Delta(f,t) = \frac{f(t+1)-f(t)}{2}$, $f_e$ is a function of the empirical time series data at time $t$, and $f_s$ is the same function of the simulated data. In essence, this equation highlights areas where the rates of change are the same magnitude, or if different, within a small distance of each other.

Across all simulation results, there are three points in the time series data which we investigate further to provide extra qualitative evidence of media mask-wearing attitudes driving opinion in the aggregate. Moreover, they are instances where partisans with different media diets received different stories about the same events, driving mask-wearing attitudes in opposite directions.

\subsubsection{New York City's Mask Mandate: 4/15/2020}

There is a noticeable spike in our simulated opinion for Republican media consumers around the date of April 15, 2020. On that day, former New York City Governor Cuomo issued an executive order ``requiring all people in New York to wear masks or face coverings in public.'' Our dataset reflects this, but perhaps not in the way that may be expected. 

Annotators picked up on strong signals coming from both Sean Hannity and Laura Ingraham's shows on \emph{Fox}. Some statements that were coded as strong pro-mask signals are listed in Table \ref{tab:4-16-statements}.

\begin{table}[h!]
    \centering
    \begin{tabular}{p{0.9\linewidth} | p{0.1\linewidth}}
        \textbf{Paragraph Text} & \textbf{Belief Rating}\\
        \hline
        \multicolumn{2}{l}{\textbf{Ingraham Angle}}\\
        \hline
        Now, we're going be examining the shifting goalposts from the meaning of flattening the curve, remember that? To whether we should wear masks in public? How can the public be confident in what they're hearing from the health officials if things keep changing? & 0.5\\
        \hline
        GOV. ANDREW CUOMO (D-NY): Wearing a mask is one of the best things that we can do. If you get on the bus, you need to wear a mask. If you get in a train, you need to wear a mask. & 0.9375\\
        \hline
        \multicolumn{2}{l}{\textbf{Hannity}}\\
        \hline
        Now, masks, gloves, temperature checks to get into any building. Wearing a mask in the building, half the workforce should continue to work remotely, from home, so you have more social distancing in every New York City building & 0.9285\\
        \hline
        HANNITY: -- just for a short period of time wearing a mask? I would. That's my -- I'm not -- I don't speak for all New Yorkers. & 0.9167\\
        \hline
        Incubations are also on the decline, that continues. The death rate appears to have flattened, and thanks to the president's numerous actions, New York never ran data ventilators, never ran out of hospital beds. Received millions and millions of masks, and gloves, and gowns, and shields from the federal stockpile. & 0.875
    \end{tabular}
    \caption{Examples from the Ingraham Angle and Hannity that were coded as pro-mask sentiment and contributed to a rise in Republican support for masks in our Single Point Model.}
    \label{tab:4-16-statements}
\end{table}

In the first case, Ingraham is attempting to point out a contradiction, or confusing messaging coming from health officials. This statement was coded neutrally, as containing one pro-mask attribute and one anti-mask attribute. Though, as we can see from inspection, it is clearly a message casting doubt on public health officials. In the early stages of the pandemic, many social scientists argued that trust in public officials would be crucial to adoption of public health measures \cite{VanBavel2020UsingResponse,Swire-Thompson2020PublicRecommendations}. These early messages reducing trust may have contributed to difficulty in adhering to public health guidelines from Republican media consumers.

The second statement, from Governor Cuomo, is part of the same segment, but coded as strongly pro-mask. This likely contributed to the rise in Republican sentiment, but may be a case of mislabeling, as the continuation of the Ingraham segment proceeds to critique Cuomo. This miscoding is an artifact of our procedure coding paragraphs independently rather than entire articles.

Yet in the second, Hannity is actually, perhaps surprisingly, advocating for mask-wearing. These quotes are not critiqued later in the segment, but are simply calls to wear a mask. These statements also inevitably contributed to the rise in opinion by Republicans during this period. This presents an example where messaging from one media diet is not monolithic, with different commentators reaching different conclusions about public health measures.



\subsubsection{Confrontation between President Trump and Los Angeles Mayor Garcetti over mask mandates}

Our data also highlights a short feud concerning mask-wearing between former President Trump and Los Angeles Mayor Garcetti in late May, 2020. Los Angeles had instated mask-wearing requirements in early April, and by late May, several instances of former President Trump flaunting mask-wearing requirements had made it into media headlines. In our Single Point Model, supportive mask-wearing opinion for those consuming a Republican media diet decreases during this time, while opinion for the Democrat media diet increases. Below, we list several article titles and content that demonstrate drivers for this difference.


\begin{table}[h!]
    \centering
    \begin{tabular}{p{0.9\linewidth} | p{0.1\linewidth}}
        \textbf{Paragraph Text} & \textbf{Belief Rating}\\
        \hline
        \multicolumn{2}{l}{\textbf{Ingraham Angle}}\\
        \hline
        Bipartisan resistance grows to draconian coronavirus lockdown orders & N/A\\
        \hline
        \multicolumn{2}{l}{\textbf{Breitbart}}\\
        \hline
        Nolte: Media Want to Focus on COVID-19 Deaths We Can't Do Much About & N/A\\
        \hline
        LA Mayor Garcetti to Trump: 'Real Men Wear Face Masks' & N/A\\
        \hline
        \multicolumn{2}{l}{\textbf{Tucker Carlson Tonight}}\\
        \hline
        The citizens who remain stuck in Los Angeles are effectively hostages of the Mayor. Garcetti has demanded that anyone who goes outside for any reason as the heat rises in LA must wear a mask. & 0.1667\\
        \hline
        \multicolumn{2}{l}{\textbf{New York Times}}\\
        \hline
        The New Face of Restaurant Hospitality Wears a Mask - The New York Times & N/A\\
        \hline
        'Coronavirus Live Updates: Fauci to Warn of ‘Needless Suffering and Death’ if States Open Too Soon' & N/A\\
        \hline
        \multicolumn{2}{l}{\textbf{Vox}}\\
        \hline
        'How masks helped Hong Kong control the coronavirus' & N/A\\
        \hline
        What Trump’s refusal to wear a mask says about masculinity in America & N/A\\
        \hline
    \end{tabular}
    \caption{Examples of news headlines and content from several stories revolving around mask-wearing orders in Los Angeles and President Trump's refusal to comply. Examples marked with ``N/A'' are titles of articles, and subsequently were not rated by annotators.}
    \label{tab:5-12-statements}
\end{table}

In response to this event, the split between messaging from the Democrat media diet and Republican media diet is evident. In the Republican media diet, there is repeated criticism of mask-wearing orders from the \emph{Ingraham Angle}, \emph{Breitbart}, and \emph{Tucker Carlson Tonight}.

whereas the \emph{New York Times} and \emph{Vox} continue to signal that wearing a mask and locking down is necessary to control the spread of the virus. The brief exchange between Mayor Garcetti and President Trump elicits specific, different responses from Democrat and Republican media: \emph{Vox} suggests that President Trump is being problematically masculine in his refusal to wear a mask, while Tucker Carlson attacks Garcetti as keeping Los Angeles residents hostage.

\subsubsection{Republican media anti-mask rhetoric after 5/21}

From our simulation and dataset, there is also a significant and sustained decrease in mask-wearing support from those consuming a Republican media diet from 5/21 onward, with a notable spike in anti-mask attitudes on 5/26 as displayed in Fig. \ref{fig:article-belief-over-time}. During this time, Democrat media consumers maintain their opinion, with a slight increase in pro-mask attitudes toward the end of the date range. 

To investigate the difference in media rhetoric that led to this difference in opinion, we examined some of the coded paragraphs from this time frame, listed in Table \ref{tab:5-21-statements}.

\begin{table}[h!]
    \centering
    \begin{tabular}{p{0.9\linewidth} | p{0.1\linewidth}}
        \textbf{Paragraph Text} & \textbf{Belief Rating}\\
        \hline
        \multicolumn{2}{l}{\textbf{Ingraham Angle}}\\
        \hline
        INGRAHAM: So, it was stunning to see him the very next day, walking around Virginia Beach, taking selfies with strangers right up against him, unmasked the very activity he just said could actually kill people. OK, so you obviously didn't believe that is what I'm saying earlier. They don't believe what they're saying. They just want to control you. & 0.25\\
        \hline
        Other challenges to masks have included that masks are an expressive form of communication. They communicate a message. In this case, I'm scared or, you know, as the Left would say, I care about you. And can the state compel us to carry that message forward? & 0.1\\
        \hline
        ARROYO: Well, whether it has use or not, Dr. Fauci a few months ago said it had very little use. But from Joe Biden it is virtue signaling. It's a way to set himself off from Trump. And in his first in-person interview in two months he sat down 12 feet apart from CNN's Dana Bash and explained why he wore the mask on Memorial Day. & 0.1\\
        \hline
        INGRAHAM: It's all political. And now we are reading stuff about how you can damage yourself by wearing a mask because then you are breathing back potentially microbes & 0.0715\\
        \hline
        \multicolumn{2}{l}{\textbf{Fox}}\\
        \hline
        WALLACE: Then there are the issue of masks, which we touched on before and we are seeing growing confrontations. For instance, in stores where customers say you've got to wear a mask and some customers say you are violating my rights. Here's an example of that. & 0.3\\
        \hline
        \multicolumn{2}{l}{\textbf{New York Times}}\\
        \hline
        Disney did not give reopening dates for the smallest of its six Florida parks, Typhoon Lagoon and Blizzard Beach, which offer water slides. Disney Vacation Club, a time-share business with 3,200 units at the resort, will reopen on June 22. Disney Springs, an adjacent 120-acre shopping mall, began to reopen on May 20. Guests have been extremely compliant, Mr. Chapek said of mask requirements at Disney Springs. & 0.9\\
        \hline
        \multicolumn{2}{l}{\textbf{Daily Kos}}\\
        \hline
        I'm not with him every day in every moment, so I don't know if he can maintain social distance, she said. I've asked everybody independently to really make sure that you're wearing a mask if you can't maintain the 6 feet. I'm assuming that in a majority of cases he's able to maintain that 6 feet distance. & 0.75 \\
    \end{tabular}
    \caption{Examples from both Democrat and Republican media diets that convey different messaging during the same time period.}
    \label{tab:5-21-statements}
\end{table}

These messages demonstrate a strong attack on masks by Republican media diet sources, particularly by Laura Ingraham during an airing of her show on 5/26. Ingraham invokes several arguments that were later found to be prevalent attitudes among partisan anti-mask discourse. Scoville et al. 2022 \cite{Scoville2022MaskReader} and Powdthavee et al. 2021 argue that face mask-wearing became a salient symbol of political identity and partisan animosity; the former from an analysis of media discourse throughout 2020, and the latter from a game-theoretic experiment. Ingraham's criticism of Virginia Governor Northam, saying that ``they'' want to control viewers, and picking out ``the Left,'' are all signals of group membership -- presumably against Democrats, whom Ingraham also associates with ``the Left.'' He et al. 2021 \cite{He2021WhyPandemic} found in Twitter discourse from January to October, 2020 that a prominent anti-mask rationale was also that they were not effective, and Taylor \& Asmundson 2021 found the same prominence of lack of efficacy as a major attitude expressed in their self-report survey of anti-mask attitudes from July to August, 2020 \cite{Taylor2021NegativeReactance}. Ingraham expresses these attitudes by citing the argument that wearing a mask makes you ``[breathe] back potentially microbes.'' Finally, Ingraham frames mask-wearing as an extension of free speech, which the government may infringe on by requiring mask-wearing. This language of individual rights is echoed in the Fox quote citing customers in stores who are asked to wear a mask arguing in return: ``you are violating my rights.'' Al-Ramahi et al. 2021 found from an analysis of Twitter data from January to October, 2020 that ``constitutional rights'' and ``freedom of choice'' dominated anti-mask discussions \cite{Al-Ramahi2021PublicData}. 

\begin{figure*}[!h]
    \centering
    \begin{subfigure}{\textwidth}
        \centering
        \includegraphics[width=\linewidth]{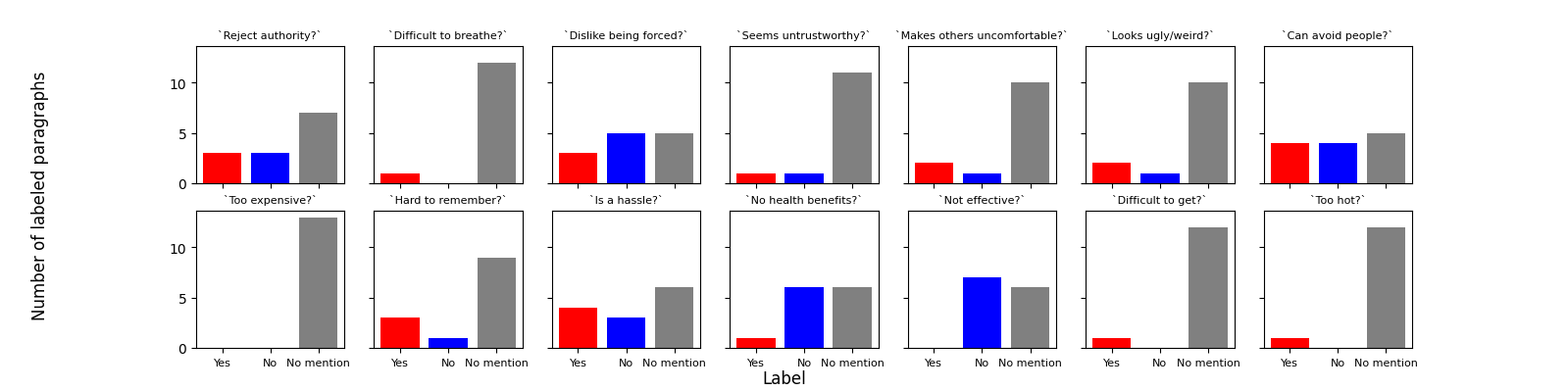}
        \caption{FMPS attitudes for media outlets in the Democrat media diet}
        \label{fig:5-22-6-8-dem-attitudes}
    \end{subfigure}
    \begin{subfigure}{1.0\textwidth}
        \centering
        \includegraphics[width=\linewidth]{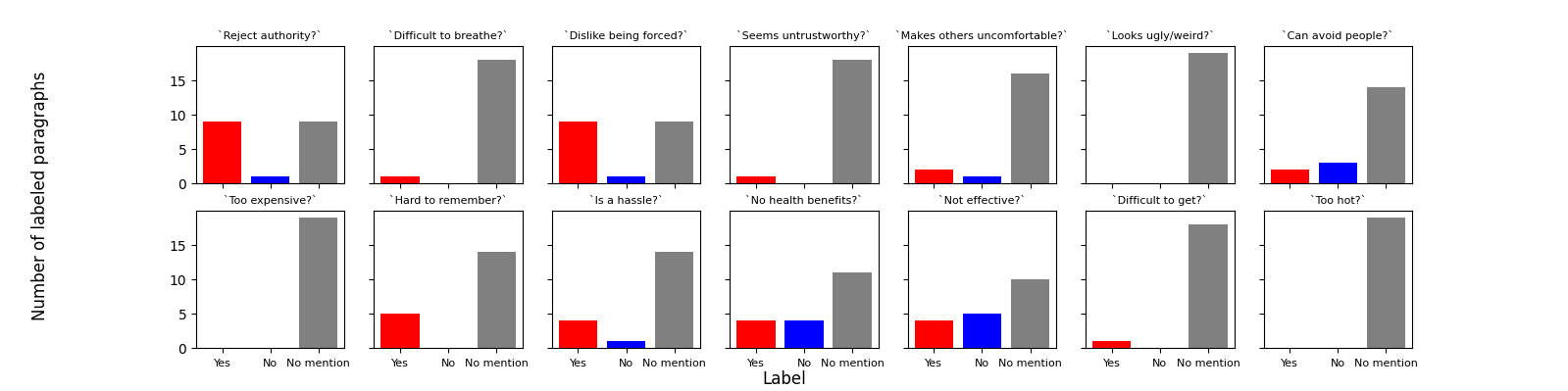}
        \caption{FMPS attitudes for media outlets in the Republican media diet}
        \label{fig:5-22-6-8-rep-attitudes}
    \end{subfigure}
    \caption{Mask-wearing attitudes along the 14 dimensions from Howard's 2020 FMPS for coded paragraphs in articles from 5/22 to 6/8, 2020 in the (a) Democrat media diet and (b) Republican media diet.}
\end{figure*}

We can also use the distribution of labels along the FMPS dimensions to examine which aspects of mask-wearing attitudes were expressed by outlets in the Democrat media diet versus the Republican media diet. For the time span from 5/22 to 6/8, these distributions are displayed in Fig. \ref{fig:5-22-6-8-dem-attitudes} (Democrat media diet) and Fig. \ref{fig:5-22-6-8-rep-attitudes} (Republican media diet).

There is a noticeable difference between the attitudes expressed by outlets in the Democrat media diet and those in the Republican media diet, but it is not entirely cut and dry. Those in the former category do express some difficulties inherent in wearing masks (e.g. that they are a hassle, people can be otherwise avoided, they are hard to remember, and that they can be uncomfortable). But those outlets also express attitudes that masks provide health benefits, are effective, and that people can be forced to wear them. From the latter category, the Republican media outlets express more attitudes that people want to prove a point against authority, that people do not like being forced to do something, and that masks are hard to remember to wear, a hassle, and may not be effective.

The attitudes within the FMPS ``independence'' dimension (that people do not like being forced to do something, and they want to prove a point against authority) are reflected in Ingraham's speech that we highlight above, as she questions the government's ability to force mask-wearing as a form of speech. Grossmann and Hopkins contend that Republican party members ``are united by a common devotion to the principle of limited government'' \cite{Grossmann2015IdeologicalPolitics}, which implies that a ``Republican'' identity may be activated by arguments in this vein. This further bolsters what we highlight above, from Ingraham's speech, as signaling partisan identity against Democrats, ``them,'' and ``the Left.''

Moreover, some of Ingraham's show's messaging casts doubt on the trustworthiness of major political officials advocating for mask wearing: Dr. Anthony Fauci and Joe Biden. Trust in government officials and institutions was cited early in the pandemic as a factor that would influence belief in health information \cite{Swire-Thompson2020PublicRecommendations,VanBavel2020UsingResponse} and came at a time when trust in journalistic institutions and science was already at historic lows \cite{Lazer2018TheEffort}. Moreover, belief in misinformation has been repeatedly argued to heavily depend on trust in information sources \cite{Pennycook2021TheNews,Ecker2022TheCorrection,Schwarz2021WhenMisinformation} (or, even, \emph{distrust} in typical information sources driving individuals to less credible ones \cite{Marwick2018WhyEffects}), and consumption of misinformation has been linked to lower trust in typical information authorities like government institutions \cite{Ognyanova2020MisinformationPower} (at least, when those institutions are not run by people from ``your team''). By depicting Dr. Fauci as hypocritical for changing his messaging about masks, or Biden as ``virtue signaling'' by wearing one, messages like those in this sample reduce trust in institutions whose guidelines were crucial for mitigating the worst of the public health crisis wrought by Covid-19.

In total, by invoking the language of individual rights, criticizing mask-wearing Democrat politicians, and questioning masks' efficacy, this batch of media speech casts doubt on mask-wearing as a legitimate public health measure. Moreover, it appeals to social psychological markers, such as partisan identity, that can make the associated anti-mask beliefs difficult to correct \cite{Oyserman2021YourAccept,Ecker2022TheCorrection}. Messages that decreased trust in public health officials and institutions also would have contributed to difficulties correcting anti-mask beliefs, as corrections from distrusted sources would not be taken as seriously \cite{Pennycook2021TheNews,Oyserman2021YourAccept}.

\section{Discussion}

We were able to create a dataset of psychologically validated Covid-19 mask-wearing attitudes in U.S. major media text, and use that data to validate our simple Single Point Model against empirical polling data. With it, we demonstrate that opinion can be correlated to the granularity of consumption of individual news stories. Our crowd-sourced annotation method also revealed differing pro- and anti-mask sentiment based on different media diets, but there were some instances where important context was missed by annotators, most likely because of experimental design. There are several directions to take this work in the future, as the data collection method can be improved to create even better sources of data that capture nuance that, for example, traditional natural language processing has trouble with. Ultimately, this dataset presents an opportunity for the validation of other models seeking to understand the dynamics of opinion update from news stories, and puts forward a new way that researchers can detect and describe complicated attitudes in news media text.

\subsection{Structuring media and polling data allows models to demonstrate fine-grained influence of media on opinion}

As previously discussed, it is difficult to demonstrate that specific media messaging drives aggregate population-level beliefs. Beliefs are often vaguely defined, so measuring their presence in media data can be highly subjective. Moreover, though media data may be publicly available, it takes a great deal of effort to identify assertions that could be argued to drive these beliefs. Other studies have, for example, shown that partisan attitudes can be argued to form from ideologically clustered media networks \cite{Benkler2018NetworkPolitics}, or from generally different media diets \cite{Bird2020IsPractices}. But empirically demonstrating that the actual attitudes expressed in media may be driving the same attitudes in the population is challenging.

By structuring news data along several dimensions of an attitude in question, identifying their presence in text, arguments about media influence can be made at a fine-grained level of detail. By combining this with polling data about the same attitudes over the same period of time, studies can put forward explanations of what happens in-between, answering the question: How do media messages lead to the aggregate outcome?

Results from our simple opinion simulation show one such case of an answer to that question. In Fig. \ref{fig:gallup-sim-comparison}, we demonstrate that there are areas in our opinion simulation results where the direction of opinion update matches that of the empirical polling data. With some of these areas corresponding to the events we highlight in our analysis, it seems that even this simple Single Point Model can be used to argue for the influence of certain news statements on opinion. With a more complete dataset, this relationship may become stronger, with more simulated areas correlating with empirical data.

Moreover, without looking in advance, our Single Point Model was able to lead us to real events that stood out as strong drivers of opinion. By simply tracking the average pro- or anti-mask sentiment in stories, according to Howard 2020's FMPS dimensions \cite{Howard2020UnderstandingScale}, and using that as a signal to change opinion, the resulting data led us to uncover two mask mandates that elicited different media responses, and then a turn to more consistent anti-mask rhetoric by right-leaning media. Moreover, our data around the first mask mandate, in New York City, presents what may be unintuitive or surprising results: that right-leaning media supported wearing a mask during that time -- which is also tracked in opinion polling as Republican and Moderate news consumers increase their willingness to wear a mask in the same period. Leaning on a stereotypical view of right-leaning media reporting of Covid-19, one may assume that a mask mandate would elicit condemnation of masks. Being able to point to the real data thus has the power to add nuance to descriptions of media response, and more accurately pinpoint instances of beneficial and harmful reporting without unnecessarily stereotyping.


Of course, our model is far more simple than how media interactions occur in reality. This dataset lays the groundwork for more complicated models of opinion formation to test their theoretical mechanisms. For example, some of the most central models to the study of opinion diffusion like that of Degroot \cite{DeGroot1974ReachingConsensus}, Hegselman-Krause \cite{Hegselmann2002OpinionSimulation}, Deffuant \cite{Deffuant2000MixingAgents}, or the Friedkin-Johnsen model \cite{Friedkin1990SocialOpinions}, can be tested for their success in matching real cases of opinion change. They may have to be slightly adapted to include the influence of media agents in the formation of subscribers' opinion, but some models have already taken up that task and can be learned from \cite{Rabb2022CognitiveNews,Rabb2023CognitivePolarization}. Additionally, more complex psychological models of political belief, like that of Van Bavel \cite{VanBavel2018TheBelief}, can be tested against the data.

\subsection{Structuring data along psychologically validated dimensions rigorously identifies attitudes in media regardless of opinion model}

Regardless of a model driving opinion outputs from media inputs, the structuring and analysis alone of large sets of media data is challenging. Some of the best studies to date rely on lists of keywords to describe topics and attitudes in media data \cite{Benkler2018NetworkPolitics,Bursztyn2020MisinformationPandemic,Koevering2022ExportingDiscussed}, as they are easy to use to analyze large collections of text. However, they sometimes lose nuance because words without their context can have different meanings. Other studies are using machine learned topic models to capture more of this nuance \cite{Pascual-Ferra2021ToxicityPandemic,Krawczyk2021QuantifyingResource,Lee2021StructuralData,Shin2022Mask-WearingElites}, but the topics identified by the algorithms also yield collections of keywords and phrases, and must be interpreted after the fact.

Our data annotation process allowed us to demonstrate a mechanism for identifying scientifically justifiable, complex attitudes in data while also capturing the way that people naturally consume media. By using Howard 2020's FMPS \cite{Howard2020UnderstandingScale}, we were able to lean on excellent psychological work to structure data in a well-defined, rigorous manner and identify Covid-19 mask-wearing attitudes in media text. Having distinct, empirically validated attitudes along orthogonal dimensions lends this process a great deal of power as it is more likely to discover attitudes that have been shown to actually exist in large populations. Studies that create lists of keywords or interpret topic models rely on one or a few researchers to determine which attitudes are extant in the data. Starting from a scientifically-grounded set of attitudes is an advantage to these methods.

Moreover, using these attitudes to hand-annotate data from a group of annotators additionally gives the data more ecological validity, as it was being read and interpreted by real people. We found several instances in our data where annotators did not agree on their interpretations of pieces of text. By employing methods from crowd-sourced data annotation \cite{Pustejovsky2012NaturalApplications}, we ensure that only annotations that were strongly agreed upon by several people drive analysis. Though there is something to be said for analyzing why differences in interpretation occur (as they could be driven by interesting sociopolitical factors), striving for annotations with consensus lends a degree of replicability to the data, and should identify less ambiguous media attitudes.

On the whole, this method seems to strike a balance between keyword detection and topic modeling, leveraging the strengths of both: it captures the meaning-driven structure of predefined topics, as in keyword analysis, but allows the human coders to capture context and linguistic nuance as the machine learning models do. 

A result of this data structure, separate from its ability to feed into a model of opinion change, is the ability to conduct a more quantitative analysis of media data that incorporates complex beliefs and ideas. Akin to work on sentiment analysis, allowing for the graded classification of pro- and anti-mask statements in the media allows for more research that can detect similar phenomena that drive important political or public health behavior. For example, our comparison between media sources in Fig. \ref{fig:media-article-belief-over-time} demonstrates at a glance that most sources had a mix of pro- and anti-mask statements, but sources like \emph{Breitbart}, \emph{Tucker Carlson Tonight}, \emph{The Ingraham Angle}, and \emph{Hannity} projected almost entirely anti-mask statements. This analysis also shows surprising results like the fact that much of Fox News' coverage is strongly pro-mask. These results give us a quantitative backing to argue that, for example, \emph{Fox}'s news coverage may differ quite dramatically from its opinion shows.

These general trends in which mask-wearing attitudes were most present in news reporting is also reflected in the results displayed in Fig. \ref{fig:labels-by-attribute}. The two attributes from Howard 2020 \cite{Howard2020UnderstandingScale} that were most prevalent, in descending order, are individuals disliking being forced to do something, and individuals rejecting authority. These attitudes strongly reflect the narratives we detected as anti-mask sentiment spiked in certain periods through typical conservative ideological disagreement with government mandates \cite{Grossmann2015IdeologicalPolitics,Al-Ramahi2021PublicData}. The same figure shows that the two most prevalent pro-mask attitudes, in descending order, are that they are effective, and have health benefits.


Being able to demonstrate the presence of often qualitatively felt trends through quantitative data is a strength of this annotation process. Moreover, it validates to some degree the utility and veracity of Howard's \cite{Howard2020UnderstandingScale} mask-wearing attitudes survey, as some of its attributes were found frequently in our news text, and their presence reflected real trends in mask-wearing sentiment in the aggregate. 

\subsection{Limitations}
\label{sec:limitations}

\subsubsection{Paragraph-based annotation limits important context}

As mentioned above, one major limitation to this work is in the annotation procedure, which presented annotators with only paragraphs of text that appear in larger news articles. In some cases, as is typical in more objective reporting, paragraphs are mostly independent from each other, and the annotation procedure is valid. However, in the cases of opinion shows, we noticed that in \emph{Tucker Carlson Tonight}, \emph{The Ingraham Angle}, and \emph{Hannity}, individual paragraphs were often pieces of larger rhetorical points -- individual paragraphs' meaning would change once contextualized with surrounding text. For example, pro-mask statements were often quoted verbatim from Democratic politicians, and then subsequently deemed hypocritical.

Moving to an annotation procedure that allows participants to take in the context of the full article is a logical next step for this work. While asking participants to read an entire article is more labor intensive, it adds important context that is needed to understand fully the pro- or anti-mask sentiment expressed in the articles.

\subsubsection{Ideological interactions with text are not tracked}

Moreover, there is inevitably an interaction of annotator ideology and political leaning with the text presented. It is argued in misinformation studies that individuals with different worldviews interact with text differently \cite{Kuo2021CriticalPolitics,Marwick2018WhyEffects}.

There is inevitably bias in our sample, as the annotators were all recruited from the same university campus, which tends to lean politically liberal. Quantifying the political leanings of annotators before they embark on the annotation task would be an important next step for this work, as then results could also be contextualized by the sociopolitical worldview of the annotators. Perhaps different worldviews would annotate text differently, and important disagreements could be detected.

\subsection{Future work}
\label{sec:future-work}

\subsubsection{Testing other opinion dynamical models on this data}

The data that we collected should be tested with other, more complicated, models of opinion dynamics. Ours was intentionally very simple; demonstrating the initial point and leaving room for other studies to improve upon our results. Using other popular opinion dynamics models like the Degroot \cite{DeGroot1974ReachingConsensus}, Hegselman-Krause \cite{Hegselmann2002OpinionSimulation}, Deffuant \cite{Deffuant2000MixingAgents}, or Friedkin-Johnsen model \cite{Friedkin1990SocialOpinions}, could add important network effects to the dynamics and improve analytical power.

Our own Cognitive Cascade model \cite{Rabb2022CognitiveNews} and its extensions to model more complex media ecosystems \cite{Rabb2023InvestigatingEcosystems} can be validated with this data. We plan to perform such a validation as a logical next step because these models incorporate much more complex dynamics than the Single Point Model used in this study: we can test how different network topologies and media consumer cognitive models of belief, driven by real media attitudes from this dataset, result in varying fits to the empirical polling data.

\subsubsection{Machine learning to label larger datasets}

The process of manually annotating with crowd-sourced annotators was arduous, and did not cover all the articles downloaded from MediaCloud for this time span. But this core of initial data can likely be used in conjunction with new semi-supervised machine learning techniques that are being used to label large datasets given the signal from an initial seed \cite{Duarte2023AClassification}. Moreover, perhaps new Large Language Models such as GPT-4 can fine-tuned to this data and then be able to detect mask-wearing sentiment in novel text. Such a model could label, for example, the rest of the articles this study collected, or any new set of articles during a different time period.

Additionally, some of the attitudes that we found expressed in our qualitative analysis of messages from certain time frames could similarly be quantified and measured. There are several aspects of political messaging that have been identified as pertinent to the belief in, and spread of, misinformation, such as source credibility and other trust cues \cite{Schwarz2021WhenMisinformation,Pennycook2021TheNews,Ecker2022TheCorrection}; drawing of in- and out-group, particularly along partisan lines \cite{Ecker2022TheCorrection,VanBavel2018TheBelief,Oyserman2021YourAccept}; or framing of issues inside of certain ideological dimensions \cite{Jost2009PoliticalAffinities,Oyserman2021YourAccept}. Being able to quantitatively identify these contributors to belief in false news at scale would aid tremendously to the study of misinformation and disinformation.

\section{Conclusion}

Our study collects a dataset of news articles from eight major U.S. media sources from April 6 to June 8, 2020, at the beginning of the Covid-19 pandemic. We recruited participants to hand-annotate news story paragraphs according to their degree of pro- or anti-mask sentiment using prompts from Howard 2020's Face Mask Perception Scale, resulting in a dataset with 2,562 high-quality annotations across 181 paragraphs from 119 distinct articles. Combined with polling data about mask-wearing attitudes during the same time frame from a Gallup poll by Bird \& Ritter, we use a simple opinion dynamic model to demonstrate that fine-grained media statements about mask wearing correlated with rises and falls in opinion polling. Our quantitative analysis of mask-wearing sentiment in our dataset further identifies real events, such as mask mandates, whose differential coverage between left- and right-leaning media sources drove opinion in different directions. While this study and analysis presents a simple initial model and example of media messaging driving opinion, it both opens the door for more complicated models of opinion dynamics to use our data to validate their models, and takes steps towards a method that can be used to annotate other sets of media data to further study its influence on public opinion.

\bibliographystyle{plain}

\section{Supplemental Materials}

\subsection{Data estimation method for Gallup opinion poll}

Data for the Gallup opinion poll described by Bird \& Ritter \cite{Bird2020IsPractices} was estimated by the following method. First, the image titled ``American Who Report Wearing Masks, Practicing Social Distancing and Advising Healthy People to Stay Home, by Party Affiliation'' was downloaded, and the section subtitled ``Wear masks'' was cropped to be the main image. A point grid system was overlaid on the graph axes using Adobe Illustrator so that one point unit corresponded to one percentage point on the y-axis of the graph -- that is, the height of the image was set to 60 points (as the graph y-axis started at 40\% and went to 100\%). The bottom of the graph was set to sit at the y-coordinate point 0. The number of days between April 6 and June 8, 2020 were calculated (64 days), and an even grid of 64 points along the x-axis was overlayed on the image.

For each intersection of the 64-day grid and the opinion time series lines, a vector line was traced (the resulting lines each had 64 points with varying y-values). For each line, each coordinate point constituting the line subsequently represented a day from April 6 to June 8 with its x coordinate, and an opinion value with its y coordinate. Because the scale was set so the top of the graph was 60 points higher than y coordinate point 0, each y coordinate on the line's value in points was equivalent to the opinion value minus 40 (because the graph started at 40\% opinion). Thus, y coordinate in each vector line was translated up 40 points to yield a value that accurately represented the true opinion value.

Such a vector line was traced and translated into data for the Republican opinion, Democrat opinion, and Independent opinion. This time series data is contained in the \emph{polling-data.csv} file contained in our data repository.

\subsection{Training questions for annotation}

Below, listed in order of presentation, are the prompts that were shown to annotators during the training section of the annotation sessions. To note, each annotator was required to label each training prompt with the same 14 FMPS dimensions as listed in Table \ref{tab:mask-coding-questions}, ask any questions they had to the researcher present conducting the annotation session, and then get permission to progress past the training section and onto annotating the actual data. Moreover, some prompts were made up -- not found in the data set -- and some were taken from the actual data (this is listed in the below Table).

\begin{table}[h!]
    \centering
    \begin{tabular}{p{0.9\linewidth} | p{0.1\linewidth}}
        \textbf{Prompt Text} & \textbf{Prompt Source}\\
        \hline
        Polling has indicated that over 60\% of mask-wearers report discomfort when wearing their mask for prolonged periods of time, including difficulty breathing and soreness behind the ears. However, those who wear masks often also report that they feel it keeps them safe and is a small burden to bear for the greater good. Indeed, mask are effective in keeping you safe from both coronavirus and other airborne illnesses. & Made up\\
        \hline
        When people wear masks, it makes them scarier. I mean, when have you seen someone with half their face covered and thought to yourself, ‘I trust that person’? Besides that, you can’t find the damn things anywhere. And we’re being forced to wear them, as if that makes people more compliant. & Made up\\
        \hline
        Scientists addressing SAGE have said that while they may not be able to stop a person catching coronavirus, there is some evidence suggesting the even homemade cloth masks could help prevent the spread of droplets that carry the virus from being released into the air, according to The Times. & Data set\\
        \hline
        A White House spokesman said that Mr. Trump and Mr. Pence had both tested negative for the virus since their exposure to the aide. But the episode raised new questions about how well-protected Mr. Trump and other top officials are as they work at the White House, typically without wearing masks. & Data set\\
        \hline
        Public health experts in the state have blamed the relaxing of social distancing for the spread, as the Arizona Republic reported. Arizona began to reopen gyms, restaurants, and other businesses in mid-May. The state has not required all individuals to wear masks in public, but workers who interact with the public are expected to have a mask. & Data set\\
    \end{tabular}
    \caption{Prompts given to annotators during the training session, along with where the prompts originated.}
    \label{tab:training-prompts}
\end{table}

\end{document}